\documentclass[twocolumn,amsmath,amssymb,nofootinbib,groupedaddress, notitlepage]{revtex4-1}

\input epsf
\usepackage{graphicx}
\usepackage{color}

\newcommand{\beq}{\begin{equation}}
\newcommand{\eeq}{\end{equation}}
\newcommand{\barr}{\begin{eqnarray}}
\newcommand{\earr}{\end{eqnarray}}

\newcommand{\rme}{\textrm{e}}

\newcommand{\Tm}{T_{\rm m}}
\newcommand{\Tr}{T_{\rm r}}
\newcommand{\nh}{n_{\rm H}}

\newcommand{\bs}{\boldsymbol}

\newcommand{\apjl}{Astrophys. J. Lett.}
\newcommand{\aap}{Astron. Astrophys.}
\newcommand{\apjs}{Astrophys. J. Suppl. Ser.}

\newcommand{\mnras}{Mon. Not. R. Astron. Soc.}

\newcommand{\memras}{Mem. R. Astr. Soc. }

\newcommand{\jcap}{J. Cosm. Astropart. Phys.}

\newcommand{\lsim}{\mathrel{\hbox{\rlap{\lower.55ex\hbox{$\sim$}} \kern-.3em \raise.4ex \hbox{$<$}}}}
\newcommand{\gsim}{\mathrel{\hbox{\rlap{\lower.55ex\hbox{$\sim$}} \kern-.3em \raise.4ex \hbox{$>$}}}}

\begin{document}
\title{Effective conductance method for the primordial recombination spectrum}

\author{Yacine Ali-Ha\"imoud} 
\email{yacine@ias.edu}
\affiliation{Institute for Advanced Study, Einstein Drive, Princeton, New Jersey 08540}

\date{\today}

\begin{abstract}
As atoms formed for the first time during primordial recombination,
they emitted bound-bound and free-bound radiation leading to
spectral distortions to the cosmic microwave background. These
distortions might become observable in the future with
high-sensitivity spectrometers, and provide a new window into physical
conditions in the early universe. The
standard multilevel atom method habitually used to compute the recombination
spectrum is computationally expensive, impeding a detailed quantitative exploration of the information contained in spectral
distortions thus far. In this work it is shown that the emissivity in optically
thin allowed transitions can be factored into a computationally expensive but
cosmology-independent part and a computationally cheap, cosmology-dependent part. The
slow part of the computation consists in pre-computing
temperature-dependent effective ``conductances'', linearly relating line
or continuum intensity to departures from Saha equilibrium of the lowest-order
excited states ($2s$ and $2p$), that can be seen as ``voltages''. The computation of
these departures from equilibrium as a function of redshift is itself
very fast, thanks to the
effective multilevel atom method introduced in an earlier work. With this
factorization, the recurring cost of a single computation of the recombination spectrum is only a
fraction of a second on a standard laptop, more than four orders of magnitude shorter than
standard computations. The spectrum from helium
recombination can be efficiently computed in an identical way, and a fast code computing the
full primordial recombination spectrum with this method will be
made publicly available soon. 

\end{abstract}

\maketitle

\section{Introduction}

A significant part of our knowledge about the universe at early times
and on large distance scales is derived from the observation and analysis of its
spatial inhomogeneities. In particular, observations of the temperature and
polarization anisotropies of the cosmic microwave background (CMB) have allowed
cosmologists to determine the geometry, contents, and initial
conditions of the universe to an exquisite level of precision (see for example Ref.~\cite{WMAP}).  

Additional, and perhaps complementary information, is hidden in the
tiny but unavoidable spectral distortions to the nearly perfectly
thermal radiation background \cite{SZ_1970, Sunyaev_Chluba_2009}. On the one hand, broad spectral distortions can be
generated by energy injection in the early universe, taking the form
of chemical potential ($\mu-$type) or Compton ($y-$type) distortions (and
more generally continuously interpolating between these two analytic
cases \cite{Khatri_Sunyaev_2012}),
depending on the redshift of energy injection. Physical mechanisms causing such energy
injections include the dissipation of small-scale acoustic waves, which
occurs in the standard cosmological picture (see
e.g.~\cite{Hu_Scott_1994}), and possibly the decay or annihilation of
dark matter into standard model particles that then deposit their
energy into the primeval plasma \cite{Hu_Silk_1993}. A vast literature
exists on the subject, and we refer the interested reader to the recent works
\cite{Khatri_Sunyaev_2012, Chluba_Khatri_2012, Pajer_Zaldarriaga_2012}
and references therein. 

On the other hand, the primordial
recombinations of helium and hydrogen lead to a few distortion
photons per atom, in the form of free-bound and bound-bound
photons. The seminal papers on primordial recombination by Peebles and
Zeldovich et al.~\cite{Peebles_1968, Zeldovich_Kurt_1969}
already evaluated the distortion due to Ly-$\alpha$ transitions and $2s-1s$ two-photon
decays. Even though this
represents a large distortion to the Wien tail of the CMB blackbody
spectrum, it lies many orders of magnitude below the cosmic infrared
background (CIB, \cite{CIB}), which renders its detection very
challenging, if not hopeless. 

In addition, exactly one free-bound
photon per hydrogen atom and two per helium atom were emitted, as
well as a few bound-bound photons per atom from transitions between highly excited
states\footnote{Throughout this paper we refer to bound-bound transitions between
  ``highly excited'' states as those for which the lower state itself
  is excited, $n' \rightarrow n \geq 2$.} \cite{CS_2006}. It was first
pointed out by Dubrovich \cite{Dubrovich_1975} that these primordial
recombination lines could be observed today as broadened features in the centimeter to decimeter
wavelength range. A few authors have since then tackled the computation
of the recombination spectrum, with various degrees of approximation
or numerical convergence \cite{Bernshtein_Dubrovich_1977,
  Lyubarski_Sunyaev_1983, Rybicki_Dellantonio_1993, Burgin_2003}, and it
is only quite recently that highly-accurate computations were carried
out by Rubi\~{n}o-Mart\'{i}n, Chluba and Sunyaev \cite{Rubino_2006,
  CS_2006, CRS_2007, Rubino_2008}. 

Thus far the study of the primordial recombination spectrum has
remained the niche domain of a few aficionados. First and foremost, the
observational prospects may seem meager, as the predicted signal in the
cm-wavelengths lies a billion times below the undistorted CMB
spectrum, far below the current best upper bounds on spectral
distortions from FIRAS \cite{FIRAS}. In addition, galactic and
extragalactic foregrounds would need to be understood and subtracted
very precisely. Finally, the machinery required to compute a recombination spectrum with the
standard multilevel method is, although conceptually not very difficult,
somewhat cumbersome to implement and too computationally expensive for a
systematic analysis of its information content. 

The reward in finding such a needle-in-a-haystack signal is, however,
potentially significant. Ref.~\cite{CS_2008} showed that the
recombination spectrum is a sensitive thermometer and baryometer. It
could also provide a clean measurement of the primordial helium
abundance, before the formation of the first stars \cite{Rubino_2008,
  Sunyaev_Chluba_2009}. Finally, pre-existing spectral distortions
could lead to a significant increase of the recombination radiation
even if the initial distortions are small in absolute value
\cite{Lyubarski_Sunyaev_1983, CS_2009_features}. The recombination
spectrum could therefore be a probe of non-standard physics such as
dark-matter annihilations \cite{Chluba_2010}.

Let us point out, in addition, that technological advances should make it possible to reach sensitivities three orders of magnitude below that of
FIRAS, corresponding to distortions at the level of $\sim 10^{-8}$
(see the proposal for the PIXIE instrument \cite{PIXIE}). Spectral
distortions from recombination are only one order of magnitude weaker
than this sensitivity (and even get to the $10^{-8}$ level around 10
GHz \cite{CS_2006}), and it is not unlikely that they will be within
reach of the next generation instruments. The feasibility of
foreground subtraction at the required level has yet to be
demonstrated, but one may hope that the spatial isotropy of the signal and its very specific spectral
features should allow us to disentangle it from foreground emission.

Before any of the truly challenging issues of instrumental sensitivity
and foreground subtraction are addressed, it seems that the first task
is to undertake a detailed quantitative study of the information
content of the recombination spectrum. In order to do so, a \emph{fast}
and accurate computational method is required, so that the
large space of cosmological parameters can be efficiently
explored\footnote{It was brought to my attention by J.~Chluba that Fendt \cite{Fendt_thesis} conducted a preliminary
  study of cosmological parameter estimation from spectral distortions, using the fast interpolation algorithm \textsc{Pico} \cite{Pico}.}. Introducing such a
fast method is the purpose of the present work.

The main task in obtaining the spectrum resides in computing the
populations of the excited states, or, more precisely, their small
departures from equilibrium with one another, since the line
and continuum emissions are proportional to the latter. High-precision
spectra require accounting for excited states up to principal quantum
number $n_{\max}$ of a few hundred, resolving the angular momentum substates. With the
standard multilevel atom method, one has to invert a large
$\frac{n_{\max}^2}2\times \frac{n_{\max}^2}2$ matrix at each timestep in order to compute the
populations of the excited states (although this matrix is
sparse due to selection rules, so only $\mathcal{O}(n_{\max}^3)$
elements are nonzero). Currently the fastest code using this method
takes about one hour for $n_{\max} = 100$ and one full day for $n_{\max} =
250$ on a standard laptop, with the computation time scaling as $t \propto n_{\max}^{3.7}$
\cite{Chluba_Vasil_2010}. 

The method that we introduce here allows us to
factorize the problem into a computationally expensive part (for which
large linear systems need to be solved) that is cosmology-independent and can
be pre-computed once and for all, and a computationally cheap part that does depend on
cosmology. This method builds on and extends the effective multilevel
atom (EMLA) method introduced in an earlier work \cite{EMLA} (hereafter AH10; see also
Refs.~\cite{Burgin_2009, Burgin_2010}); it is, however, not a trivial
extension, since the EMLA method is designed for computing the free electron
fraction $x_e(z)$ and essentially collapses all the transitions between excited states into effective transition rates
into and out of $2s$ and $2p$. In this process, information not
directly necessary to the evolution of the free electron fraction is
lost, whereas our present goal is to go beyond $x_e(z)$ and compute
the full recombination spectrum.

\begin{figure}
\includegraphics[width = 85 mm]{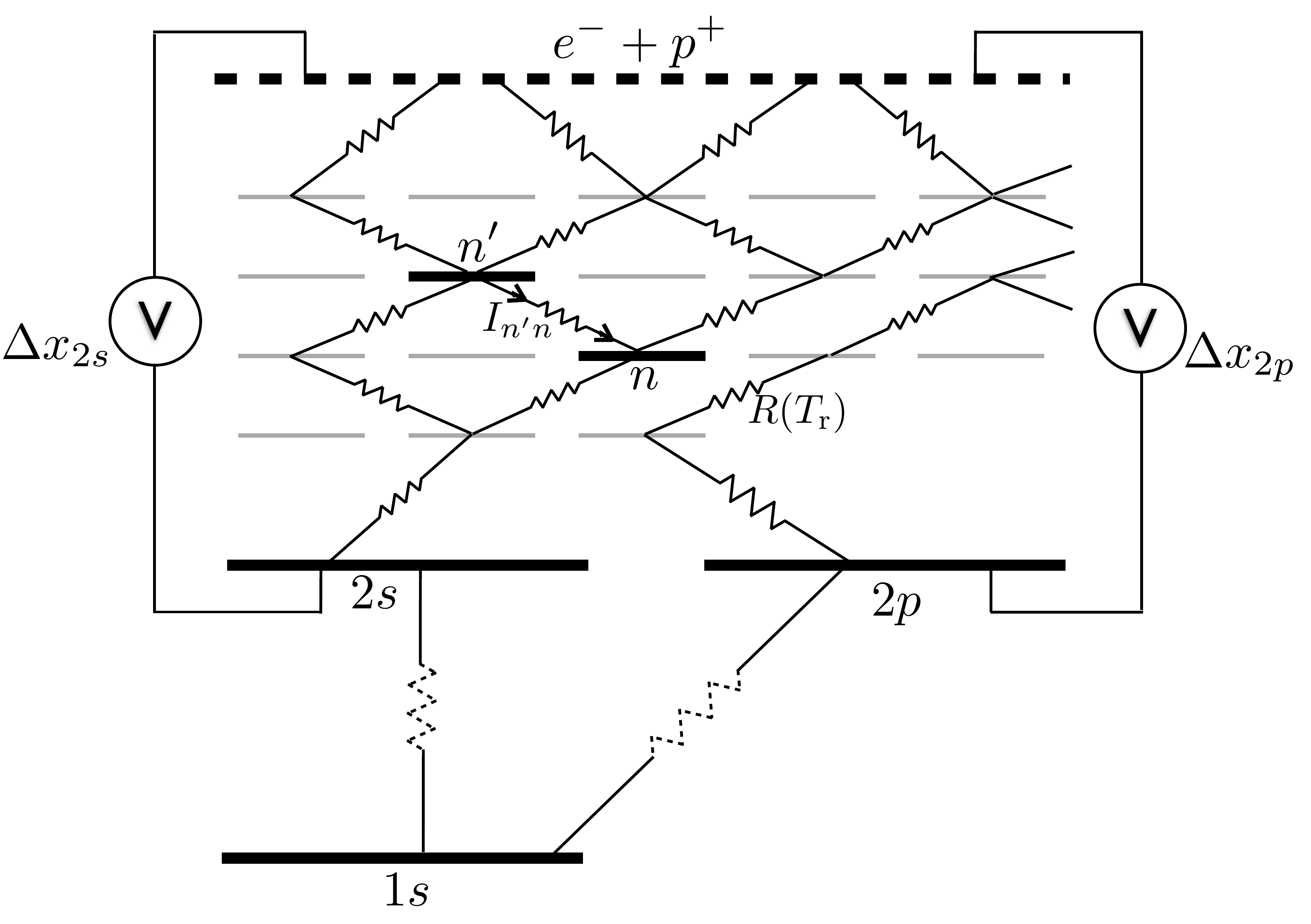}
\caption{Schematic representation of the main idea of this paper,
  highlighting the circuit analogy of Hirata \cite{Chris_blackboard}. All transitions between excited
states are mediated by blackbody photons and their rates (and corresponding
resistances $R(\Tr)$ shown as solid resistor symbols) only depend on the
blackbody temperature. The linearity of the steady-state rate equations
(Kirchhoff's laws in the circuit analogy) ensures that the current
$I_{n'n}$ in any transition between excited states is linear in the
departures from Saha equilibrium $\Delta x_{2s}$ and $\Delta x_{2p}$,
which play the role of externally imposed voltages. The proportionality coefficient is a
function of temperature only. The values of $\Delta x_{2s}$ and $\Delta x_{2p}$ are determined from the
balance between the net downward current from the highly excited
states (obtained through the effective rates defined in AH10) and the
non-thermal transition rates to the ground state (represented by
dashed resistor symbols). We have not represented all allowed
transitions in order to keep the graph readable.} \label{fig:schema} 
\end{figure}

We shall lay down our method in detail in the remainder of this paper,
but the main idea can be intuitively understood if one pictures the system of radiatively connected excited
levels as a circuit, where the currents are the line intensities, the
voltages are the departures of the excited states populations from Saha equilibrium, and the
conductances are the transition rates (this insightful analogy is due
to Chris Hirata \cite{Chris_blackboard}). The linearity of Kirchhoff's laws (the steady-state
rate equations for the excited state) ensures that the
``current'' in any transition is proportional to the outer
``voltages'', i.e. the departures from Saha equilibrium of the $2s$
and $2p$ states. The proportionality coefficients, which we shall call
effective conductances, moreover only depend on the temperature of the
ambient blackbody radiation that is nearly undistorted at the relevant
frequencies (this can in principle be generalized to include simple
parameterizations of the ambient spectrum, as well as collisional transitions). Once the effective
conductances are pre-computed as a function of transition energy and
temperature, the recombination spectrum can be computed very
efficiently for any cosmology, by using the EMLA method to evaluate
the recombination history and the outer ``voltages'' as a function of
redshift. This method is very similar in spirit to the widely used
line-of-sight integration method for CMB anisotropies
\cite{Seljak_Zaldarriaga_1996}, which factorizes the computation of
the CMB power-spectrum
into a geometric, cosmology-independent part and a cosmology-dependent
but multipole-independent source term. We illustrate our method
graphically in Fig.~\ref{fig:schema}.

This paper is organized as follows. In Sec.~\ref{sec:equations}, we
write down the general equations that need to be solved for the
computation of the recombination spectrum. The effective conductance
method is described in Sec.~\ref{sec:conductance}. We discuss our
numerical implementation and results in Sec.~\ref{sec:results} and give
our conclusions and future research directions in Sec.~\ref{sec:conclusion}.

\section{General equations} \label{sec:equations}

\subsection{Notation}

We denote by $\nh$ the total number density of hydrogen in all its forms
(ionized and neutral), $x_e$ the ratio of the free-electron abundance
to the total hydrogen abundance, $x_p$ the fraction of ionized
hydrogen, and $x_{nl}$ (or in some cases $X_{nl}$) the fractional abundance of neutral hydrogen in
the excited state of principal quantum number $n$ and angular
momentum number $l$. The matter and radiation
temperatures are denoted by $\Tm$ and $\Tr$, respectively. We denote emissivities
by $j_{\nu}$ (with units of energy per unit time per frequency interval per unit
volume per unit solid angle) and specific intensities by $I_{\nu}$
(with units of energy per unit time per frequency interval per unit
area per unit solid angle). All our derivations are for hydrogen atoms
but the generalization to helium is straightforward. We only consider
radiative transitions here and neglect the effect of
collisions. 

\subsection{Bound-bound emission from transitions between excited states}

The emissivity due to bound-bound transitions between excited states
is given by
\barr
j_{\rm bb}(\nu) = \nh \frac{h \nu}{4 \pi} \times ~~~~~~~~~~~~~~~~~~~~~~~~~~~~~~~~~~~~~~~~~~~\nonumber\\
\sum_{2 \leq n<n'}\sum_{l,l'}
\left[x_{n'l'} R_{n'l'\rightarrow nl} - x_{nl} R_{nl \rightarrow
    n'l'}\right] \delta(\nu - \nu_{n'n}),
\earr
where $\nu_{n'n}$ is the frequency of the $n'l' \rightarrow nl$
transitions and $R_{nl\rightarrow n'l'}$ represents the radiative
transition rate from $nl$ to $n'l'$. The recombination process
adds at most a few photons per atom, and the transitions between
excited states are mostly below the peak of the blackbody spectrum
(except for Balmer transitions, but their energy is only a few
times above the blackbody peak, where there is still a very large
number of thermal photons per hydrogen atom). As a consequence, the
radiation field mediating the transitions is, to $\sim 10^{-8}$ accuracy, a blackbody at temperature $\Tr$. One can check the validity
of this assumption a posteriori once the distortions are computed (see
for example Fig.~2 of Ref.~\cite{Rubino_2006}). Note, however, that
small $y$-distortions to the blackbody spectrum (at the level of $y \sim 10^{-6}$) may significantly
enhance the hydrogen and helium line emission
\cite{Lyubarski_Sunyaev_1983, CS_2009_features}. We defer the study of the effect of such
pre-existing distortions on the recombination spectrum to future work,
and shall here assume that transitions between excited states are
mediated by thermal photons only.

With this assumption, the radiative transition rates satisfy the detailed balance relations,
\beq
R_{n'l' \rightarrow nl} = \frac{q_{nl}}{q_{n'l'}} R_{nl \rightarrow
  n'l'}, \label{eq:dbR}
\eeq
where we have defined
\beq
q_{nl} \equiv (2 l +1) \rme^{-E_n/k\Tr},
\eeq
where $E_n \equiv - 13.6~ \textrm{eV}/n^2$ is the (negative) energy of the
bound states with principal quantum number $n$.

We see that if the excited states were in Boltzmann equilibrium, so
that $x_{n'l'}/x_{nl} = q_{n'l'}/q_{nl}$, the net emissivity would
vanish. The emissivity therefore scales linearly with the small departures from
equilibrium. To make this apparent, let us define the departures from
Saha equilibrium at temperature $\Tr$ with the free electron and protons,
\beq
\Delta x_{nl} \equiv x_{nl} - \frac{q_{nl}}{q_e} \nh x_e x_p, \label{eq:Dxnl-def}
\eeq
where we have defined
\beq
q_e \equiv \left(\frac{2 \pi m_e k \Tr}{h^2}\right)^{3/2},
\eeq
where $m_e$ is the reduced mass of the electron-proton system. We can
now rewrite the bound-bound emissivity in the following form:
\barr
j_{\rm bb}(\nu) = \nh \frac{h \nu}{4 \pi} \times ~~~~~~~~~~~~~~~~~~~~~~~~~~~~~~~~~~~~~~~~~~~\nonumber\\
\sum_{2 \leq n<n'}\sum_{l,l'} \left[\frac{q_{nl}}{q_{n'l'}}\Delta x_{n'l'}  - \Delta x_{nl}\right] R_{nl \rightarrow
    n'l'} \delta(\nu - \nu_{n'n}),~~
\earr
where the Saha-equilibrium pieces have cancelled out.

\subsection{Free-bound emission}

The emissivity due to free-bound transitions to excited states is
given by
\beq
j_{\rm fb}(\nu) = \nh \frac{h \nu}{4 \pi} \sum_{n \geq 2}\sum_{l<n}
\left[\nh x_e x_p \frac{d \alpha_{nl}}{d \nu} - x_{nl} \frac{d
    \beta_{nl}}{d \nu} \right],
\eeq
where $d \alpha_{nl}/d \nu$ is the differential recombination
coefficient per frequency interval of the emitted photon and $d
\beta_{nl}/d\nu$ is the differential photoionization rate per
frequency interval of the ionizing photon. Recombinations of the
thermal electrons and protons to the excited
states are mediated by blackbody photons; as a consequence, $d \alpha_{nl}/d \nu$ and $d
\beta_{nl}/d\nu$ are related through the relation (see for example
Eq.~(2) of AH10):
\barr
\frac{d \alpha_{nl}}{d \nu} &=& \left(\frac{\Tr}{\Tm}\right)^{3/2}
\exp\left[(h \nu + E_n)\left(\frac1{k\Tr} -
    \frac1{k\Tm}\right)\right]\nonumber\\
&\times& \frac{q_{nl}}{q_e} \frac{d \beta_{nl}}{d \nu}.
\earr
We have purposefully made the ratio and difference of the matter and
radiation temperatures appear in this expression. At high redshifts
when the recombination spectrum is emitted, the matter temperature is
locked to the radiation temperature by Compton heating (see for
example Ref.~\cite{recfast}). Computing the coupled evolution of the ionization history and
matter temperature with \textsc{HyRec}\footnote{http://www.sns.ias.edu/$\sim$yacine/hyrec/hyrec.html} \cite{Hyrec}, we find that the fractional
difference between the two temperatures is less than $10^{-5}$ for $z \gtrsim 1100$ and less than
$10^{-3}$ for $z \gtrsim 800$. We can therefore Taylor-expand the above expression in the small
parameter $\Delta T/T \equiv (1 - \Tm/\Tr)$ (note that with this
convention $\Delta T > 0$ for $\Tm < \Tr$) and obtain
\barr
\frac{d \alpha_{nl}}{d \nu} &\approx& \frac{q_{nl}}{q_e} \frac{d
  \beta_{nl}}{d \nu} \left[1 + \left(\frac32 - \frac{h (\nu - \nu_{cn})}{k \Tr}\right) \frac{\Delta T}{T}\right]\nonumber\\
&\equiv& \frac{q_{nl}}{q_e} \frac{d
  \beta_{nl}}{d \nu} \left[1 + \gamma_{n}(\nu) \frac{\Delta T}{T}\right],
\earr 
where $\nu_{cn} \equiv -E_n/h$ is the photoionization threshold from
the $n$-th shell and the second expression defines the dimensionless parameter
$\gamma_n(\nu)$. 
Again, we see that the free-bound emissivity would vanish if the
excited states were in Saha equilibrium with the ionized plasma and if
the matter and radiation temperature were identical. We can rewrite
the free-bound emissivity in terms of small departures from equilibrium
as follows:
\barr
&&j_{\rm fb}(\nu) = \nh \frac{h \nu}{4 \pi} \times\nonumber\\
&&\sum_{n \geq 2}\sum_{l<n} \left[\frac{q_{nl}}{q_e} \nh x_e x_p  \gamma_n(\nu) \frac{\Delta T}{T} - \Delta x_{nl}\right] \frac{d
    \beta_{nl}}{d \nu}.
\earr

\subsection{Radiative transfer equation}

Strictly speaking, the computation of the specific intensity requires
the knowledge of not only the emissivity but also the absorption coefficient
\cite{Rybicki_Lightman}. However, the Sobolev optical depths of most bound-bound
transitions are much lower than unity \cite{Hirata_2008,
  Grin_Hirata_10}, perhaps with the exception of the very high-$n$
transitions ($n \gtrsim 300$), as can be seen from extrapolating Fig.~11
of Ref.~\cite{CRS_2007}. It is possible that a proper accounting of the nonzero optical depth in very high-$n$
transitions could lead to small modifications in the low-frequency part of
the spectrum ($\nu \lesssim 0.1$ GHz); however, in that frequency range other effects that we
are neglecting significantly affect the spectrum too, such as free-free absorption
\cite{CRS_2007} and collisional transitions \cite{Chluba_Vasil_2010}. We shall therefore assume the optically-thin limit for
all bound-bound transitions between excited states, as well as
free-bound transitions.

The radiative transfer equation in the optically thin regime in an
expanding homogeneous universe takes the simple form
\beq
\frac{d}{dt}\left(\frac{I_{\nu}}{\nu^3}\right) \equiv
\left[\frac{\partial}{\partial
    t}\left(\frac{I_{\nu}}{\nu^3}\right) - H \nu \frac{\partial}{\partial
    \nu}\left(\frac{I_{\nu}}{\nu^3}\right) \right] = c\frac{j_{\nu}}{\nu^3}.
\eeq
In between resonances (where $j_{\nu} = 0$), the quantity $I_{\nu}/\nu^3$ is conserved along
a photon trajectory, so that
\beq
I_{\nu}(z) = \left(\frac{\nu}{\nu'}\right)^3 I_{\nu'}(z'),\label{eq:free-redshifting}
\eeq
where
\beq
1 + z' = \frac{\nu'}{\nu}(1 + z). \label{eq:redshift}
\eeq
In the vicinity of a resonance line, $j_{\nu} = J_0 \delta(\nu -
\nu_0)$, the solution to the radiative transfer equation is 
\beq
I_{\nu_0}^- = I_{\nu_0}^+ + \frac{c J_0}{H \nu_0}, \label{eq:jump}
\eeq
where $I_{\nu_0}^+$ and $I_{\nu_0}^-$ are the specific intensities at
the blue and red sides of the resonant line, respectively.

The general solution of the radiative transfer equation can be written
in the following integral form
\barr
I_{\nu}(z) &=& c \int_{\nu}^{\infty} d \nu' \left(\frac{1+z}{1+z'}\right)^3
\frac{j_{\nu'}}{H \nu'}\Big{|}_{z'} \nonumber\\
&=& c ~\nh(z) \int_{\nu}^{\infty} \frac{d \nu'}{H \nu'} \frac{j_{\nu'}}{\nh}\Big{|}_{z'}, \label{eq:Inu-nu'}
\earr
where the redshift $z'$ and frequency $\nu'$ are related through
Eq.~(\ref{eq:redshift}) and we used the fact that $\nh(z) \propto (1+z)^3$.

\section{The effective conductance method} \label{sec:conductance}

\subsection{Populations of the excited states}
In order to obtain the bound-bound and free-bound emissivities, we see
that we need to evaluate the populations of the excited states, or,
more precisely, their small departures from Saha equilibrium with the
plasma.

The populations of excited states can be obtained to very high accuracy
in the steady-state approximation, because of the large ratio of
internal transition rates to the overall recombination rate. This
assumption was checked explicitly in Ref.~\cite{Chluba_Vasil_2010} and found to be
extremely accurate, for the computation of both the recombination
history and the recombination spectrum.

Following AH10, we separate the excited states in ``interior'' states,
only connected radiatively to other excited states and the continuum, and ``interface''
states, essentially $2s$ and $2p$ (and potentially any additional ``weak interface'' states, such as $3s,
3p, 3d$...), which are radiatively connected to the ground state. We denote the populations of the former by a capital
$X_K$, where $K$ stands for both quantum numbers of the state, and
those of the
latter by $x_i$, where $i = 2s, 2p$ (and $3s, 3p, 3d$... if needed). 

The steady-state rate equation for the population of the interior
state $K$ is
\barr
0 \approx \dot{X}_K &=& \nh x_e x_p \alpha_K +  \sum_i x_i R_{i
  \rightarrow K} \nonumber\\
&+&\sum_{L \neq K} X_L R_{L \rightarrow K} - X_K \Gamma_K, \label{eq:ssXK}
\earr
where $\alpha_K(\Tm, \Tr)$ is the total recombination coefficient to the state
$K$ (accounting for stimulated recombinations) and 
\beq
\Gamma_K \equiv \beta_K + \sum_{L \neq K} R_{K \rightarrow L} +
  \sum_i R_{K \rightarrow i}
\eeq
is the total rate of transitions out of the state $K$, where $\beta_K$
is the total photoionization rate from $K$. Note that here again we
have assumed that the Sobolev escape probability is unity in all
transitions (or that the Sobolev optical depth is zero). If this were not the case the net transition rates would
depend non-linearly on the state populations, which would
significantly complicate matters.

Provided the transitions between excited states are mediated by
blackbody photons, the transition rates satisfy the detailed
balance relation Eq.~(\ref{eq:dbR}). The recombination coefficients
and photoionization rates are related through
\beq
\alpha_K(\Tm = \Tr) = \frac{q_K}{q_e} \beta_K(\Tr).
\eeq
We can now rewrite the system
in terms of the small departures from Saha equilibrium with the
continuum, and to linear order in $\Delta T/T$:
\barr
0 &=& -\nh x_e x_p \Delta T \frac{\partial \alpha_K}{\partial \Tm}\Big{|}_{\Tm = \Tr} +  \sum_i \Delta x_i R_{i
  \rightarrow K} \nonumber\\
&+&\sum_{L \neq K} \Delta X_L R_{L \rightarrow K} - \Delta X_K \Gamma_K. \label{eq:ssDXK}
\earr
In the standard multilevel atom method, the large system (\ref{eq:ssDXK}) is solved
at every timestep, which makes the computation very slow.

Following AH10, we define the matrix $\bs M$ of elements
\beq
M_{KL} \equiv \Gamma_K \delta_{KL} - (1 - \delta_{KL}) R_{K\rightarrow L}.
\eeq
We also define the vector $\Delta \bs{X}$ of elements $\Delta X_K$ and
the source vector $\Delta \bs{S}$ of elements
\beq
\Delta S_K \equiv -\nh x_e x_p \Delta T \frac{\partial \alpha_K}{\partial \Tm}\Big{|}_{\Tm = \Tr}
+  \sum_i \Delta x_i R_{i
  \rightarrow K}. 
\eeq
The system (\ref{eq:ssDXK}) can be rewritten in compact matrix form as
\beq
\bs{M}^{\rm T} (\Delta \bs{X}) = \Delta \bs{S},
\eeq
where $\bs{M}^{\rm T}$ is the transpose of $\bs{M}$.
We showed in AH10 that the matrix $\bs{M}$ is nonsingular, and this
system has the formal solution
\beq
\Delta \bs{X} = (\bs{M}^{\rm T})^{-1}(\Delta \bs{S}) =  (\bs{M}^{-1})^{\rm T}(\Delta \bs{S}), 
\eeq
i.e., explicitly,
\barr
\Delta X_K &=& \sum_L (\bs{M}^{-1})_{LK} \Delta S_L\nonumber\\
&=& -\nh x_e x_p \Delta T \sum_L (\bs{M}^{-1})_{LK} \frac{\partial
  \alpha_L}{\partial \Tm}\Big{|}_{\Tm = \Tr}\nonumber\\
 &+& \sum_i \Delta x_i
\sum_L (\bs{M}^{-1})_{LK}  R_{i \rightarrow L}. \label{eq:DXK-sol}
\earr
We showed in AH10 that detailed balance relations between radiative
transition rates ensure that
\beq
(\bs{M}^{-1})_{LK} = \frac{q_K}{q_L} (\bs{M}^{-1})_{KL}.
\eeq
Using the detailed balance relation for $R_{i \rightarrow L}$, we may
  rewrite the last term of Eq.~(\ref{eq:DXK-sol}) as
\barr
\sum_L (\bs{M}^{-1})_{LK}  R_{i \rightarrow L} &=& \frac{q_K}{q_i}\sum_L
(\bs{M}^{-1})_{KL}  R_{L \rightarrow i} \nonumber\\
&\equiv& \frac{q_K}{q_i} P_K^i, \label{eq:PKi}
\earr
where $P_K^i$ is the probability that an excited atom initially in the
interior state $K$ eventually reaches the interface state $i$ (after
possibly many transitions in the interior), the
complementary events being to eventually reach one of the other
interface states or to be photoionized. The probabilities $P_K^i$ were
defined in AH10, where they were an intermediate step to compute the
effective recombination coefficients to and effective transition rates
between the interface states. In the last line of Eq.~(\ref{eq:PKi}),
we have used the formal solution of the defining equation for the
probabilities $P_K^i$.

Let us now deal with the first term of Eq.~(\ref{eq:DXK-sol}). We
define the dimensionless coefficients
\beq
\gamma_K \equiv - \frac{\partial
 \log \alpha_K}{\partial \log \Tm}\Big{|}_{\Tm = \Tr}.
\eeq
Using again detailed balance relations, we obtain
\barr
\sum_L (\bs{M}^{-1})_{LK} \frac{\partial
  \alpha_L}{\partial \Tm}\Big{|}_{\Tm = \Tr} &=& - \frac1{\Tr}
\frac{q_K}{q_e}\sum_L (\bs{M}^{-1})_{KL} \beta_L \gamma_L\nonumber\\
&\equiv& - \frac1{\Tr}
\frac{q_K}{q_e} \tilde{P}_K^e, \label{eq:PKe}
\earr
where the last equality \emph{defines} the dimensionless coefficient
$\tilde{P}_K^e$. If all the coefficients $\gamma_L$ were equal to
unity, then we would have $\tilde{P}_K^e = P_K^e$, the probability
that an excited atom initially in the interior state $K$ eventually
gets photoionized before reaching an interface state. In general,
however, $\gamma_L \neq 1$ (but is in general positive and of order unity), so the numbers
$\tilde{P}_K^e$ do not have a clear physical significance but are
numerically of the same order as the $P_K^e$. 

We therefore end up with the following compact 
expression for the departures from Saha equilibrium:
\beq
\Delta X_K = \frac{q_K}{q_e}
\tilde{P}_K^e \nh x_e x_p \frac{\Delta T}{T} + \sum_i \frac{q_K}{q_i}
P_K^i \Delta x_i.\label{eq:pop-formal}
\eeq
Looking more closely at equation (\ref{eq:pop-formal}), we see that
the departure from Saha equilibrium of any excited state is
proportional to the difference between matter and radiation
temperature (so long as this difference is small) times $\nh x_e x_p$, and to the
departures from Saha equilibrium of the small set of interface
states. The proportionality coefficients are functions of radiation temperature
only. If we \emph{define} $P_j^i \equiv \delta_{ij}$ and $\tilde{P}_i^e
\equiv 0$ for interface states $i, j$, we can write a general equation valid for any excited state
(including the interface states) in the form
\beq
\Delta x_{nl} = \frac{q_{nl}}{q_e}
\tilde{P}_{nl}^e \nh x_e x_p \frac{\Delta T}{T} + \sum_i \frac{q_{nl}}{q_i}
P_{nl}^i \Delta x_i.\label{eq:pop-final}
\eeq

\subsection{Effective conductances}

We can now rewrite the net decay rate in the $n' \rightarrow n$
transitions (with $n < n'$), per hydrogen atom, in the form
\barr
\sum_{l,l'} \left[\frac{q_{nl}}{q_{n'l'}}\Delta x_{n'l'}  - \Delta x_{nl}\right] R_{nl \rightarrow
    n'l'}\nonumber\\
 = \mathcal{G}_{n'n}^e \nh x_e x_p \frac{\Delta T}{T} -
  \sum_i \mathcal{G}_{n'n}^i \Delta x_i, \label{eq:bb-conductance}
\earr
where we have defined the coefficients
\barr
\mathcal{G}_{n'n}^e(\Tr) &\equiv& \sum_{l,l'}
\frac{q_{nl}}{q_e}\left[\tilde{P}_{n'l'}^e - \tilde{P}_{nl}^e\right] R_{nl \rightarrow
    n'l'},\\
\mathcal{G}_{n'n}^i(\Tr) &\equiv& \sum_{l,l'}\frac{q_{nl}}{q_i}\left[P_{nl}^i - P_{n'l'}^i\right] R_{nl \rightarrow
    n'l'}. \label{eq:Gnpn}
\earr
Note the minus sign in Eq.~(\ref{eq:bb-conductance}): we have chosen this
convention because the
excited states are in general under-populated with respect to Saha
equilibrium, and the coefficients $\mathcal{G}_{n'n}$ defined in
Eq.~(\ref{eq:Gnpn}) are positive (in general the probability of
reaching interface states decreases as $n$ increases).

If one thinks of the departures from Saha equilibrium $\Delta x_i$ as
voltages and of the net decay rate in the $n'\rightarrow n$
transitions as a current, then the coefficients $\mathcal{G}_{n'n}^i$
can be thought of as effective conductances linearly relating the two. A similar
analogy can be made for the first term: there, the voltage is the fractional
temperature difference $\Delta T/T$ and the conductance would be $\nh
x_e x_p \mathcal{G}^e_{n'n}$. Note that the $\mathcal{G}^e$ have units of cm$^3$
s$^{-1}$ whereas the $\mathcal{G}^i$ have units of s$^{-1}$.

Similarly, the net free-bound decay rate per unit frequency can be
rewritten as
\barr
\sum_{n \geq 2}\sum_{l<n} \left[\frac{q_{nl}}{q_e} \nh x_e x_p  \gamma_n(\nu) \frac{\Delta T}{T} - \Delta x_{nl}\right] \frac{d
    \beta_{nl}}{d \nu} \nonumber\\
= \frac{d \mathcal{G}^e_{\rm fb}}{d \nu} \nh x_e x_p \frac{\Delta
  T}{T} - \sum_i\frac{d \mathcal{G}^i_{\rm fb}}{d \nu} \Delta x_i, 
\earr
where we have defined the differential effective conductances
\barr
\frac{d \mathcal{G}^e_{\rm fb}}{d \nu} &\equiv&  \sum_{n \geq
  2}\sum_{l<n} \frac{q_{nl}}{q_e} \left[\gamma_n(\nu) - \tilde{P}_{nl}^e
  \right]\frac{d \beta_{nl}}{d \nu},\\
\frac{d \mathcal{G}^i_{\rm fb}}{d \nu} &\equiv& \sum_{n \geq
  2}\sum_{l<n} \frac{q_{nl}}{q_i} P_{nl}^i\frac{d \beta_{nl}}{d \nu}.
\earr
Here again we have defined the effective conductance $d
\mathcal{G}^i_{\rm fb}/d \nu$ with a positive sign, such that the
current is positive when the interface states are underpopulated with
respect to Saha equilibrium.

\subsection{Bound-bound and free-bound emissivities}

The expressions for the emissivities can now be rewritten in terms of
the effective conductances:
\barr
&&j_{\rm bb}(\nu) = \nh \frac{h \nu}{4 \pi} \times \nonumber\\
&&\sum_{n< n'} \left[ \mathcal{G}_{n'n}^e \nh x_e x_p \frac{\Delta T}{T} -
  \sum_i \mathcal{G}_{n'n}^i \Delta x_i\right] \delta(\nu - \nu_{n'n}),\\
&& j_{\rm fb}(\nu) = \nh \frac{h \nu}{4 \pi} \left[
   \frac{d\mathcal{G}_{\rm fb}^e}{d \nu} \nh x_e x_p \frac{\Delta T}{T} -
  \sum_i \frac{d \mathcal{G}_{\rm fb}^i}{d \nu} \Delta x_i\right].
\earr
We can rewrite the total emissivity formally as
\barr
&&j_{\nu} \equiv j_{\rm bb}(\nu) + j_{\rm fb}(\nu) = \nh \frac{h \nu}{4 \pi} \times \nonumber\\
&&\left[\frac{d\mathcal{G}^e}{d \nu}(\nu, \Tr) ~\nh x_e x_p \frac{\Delta T}{T} -
  \sum_i \frac{d \mathcal{G}_{}^i}{d \nu}(\nu, \Tr) \Delta x_i
\right], \label{eq:master-eq}
\earr
where
\beq
\frac{d\mathcal{G}^{e,i}}{d \nu} \equiv \sum_{n< n'} \mathcal{G}_{n'n}^{e,i}
\delta(\nu - \nu_{n'n}) + \frac{d\mathcal{G}_{\rm fb}^{e,i}}{d \nu}.
\eeq
Equation (\ref{eq:master-eq}), along with the definitions of effective
conductances given above, constitutes the fundamental result of this
paper. What it means is that the emissivity can
be factored into a cosmology-independent part, embodied in the effective conductances, that is
computationally expensive but can be pretabulated as a function of temperature, and
a simple cosmology-dependent part, entering through the departures
from Saha equilibrium of $2s$ and $2p$ and temperature differences, which are
straightforward to compute for each particular cosmology with the
effective multilevel atom method developed in AH10.

Before proceeding further, let us point out
that even though we have Taylor-expanded our expressions in the small
difference between matter and radiation temperature, this is not
required. One can very easily obtain more general expressions for
arbitrary $\Delta T/T$, in which the coefficient $d \mathcal{G}^e/d \nu$
would become a function of radiation \emph{and} matter temperatures. We leave it as an exercice for the interested reader to derive the exact expressions.

\section{Implementation and results} \label{sec:results}

\subsection{Computation of effective conductances}

The computation of the effective conductances requires solving large
(formally infinite) linear systems. One must impose some truncation
criterion to make the system finite and tractable. Following previous
works, we simply ignore all excited levels above some cutoff value of
the principal quantum number\footnote{Other truncation schemes
could be imagined, such as assuming that the excited states above some
threshold are in Saha equilibrium with the continuum; in practice, as
long as a clear convergence is exhibited with increasing $n_{\max}$,
we need not worry about the detailed truncation prescription.} $n_{\max}$. 

We tabulated the effective conductances on a grid of temperatures for
several values of $n_{\max}$ ranging from 60 to 500. Matrix elements
for bound-bound transitions were computed as in Ref.~\cite{Hey_2006} and
those for free-bound transitions as in Ref.~\cite{Burgess_1965}. Equations (\ref{eq:PKi}) and (\ref{eq:PKe}) for the
probabilities $P_K^i(\Tr)$ and the dimensionless numbers
$\tilde{P}_K^e(\Tr)$ represent the time consuming part of the problem,
as they require solving large (of order $n_{\max}^2/2 \times
n_{\max}^2/2$) matrix equations. Due to selection rules for radiative
transitions, these large matrices are very sparse, and only have of
order $n_{\max}^3$ nonvanishing elements. We can therefore use a
sparse matrix technique identical to the one introduced in
Ref.~\cite{Grin_Hirata_10}. 

\begin{figure*}
\includegraphics[width = 130 mm]{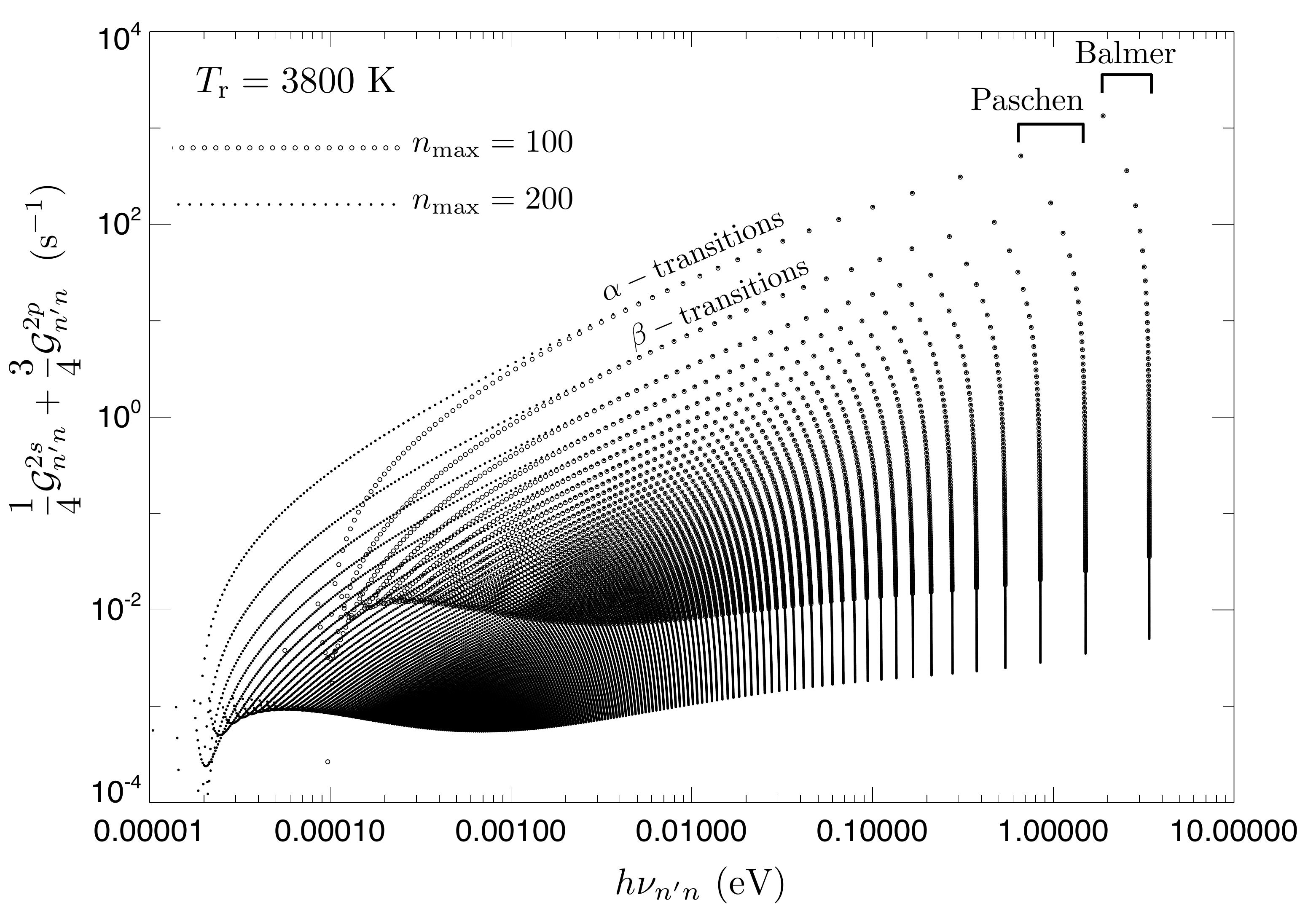}
\includegraphics[width = 130 mm]{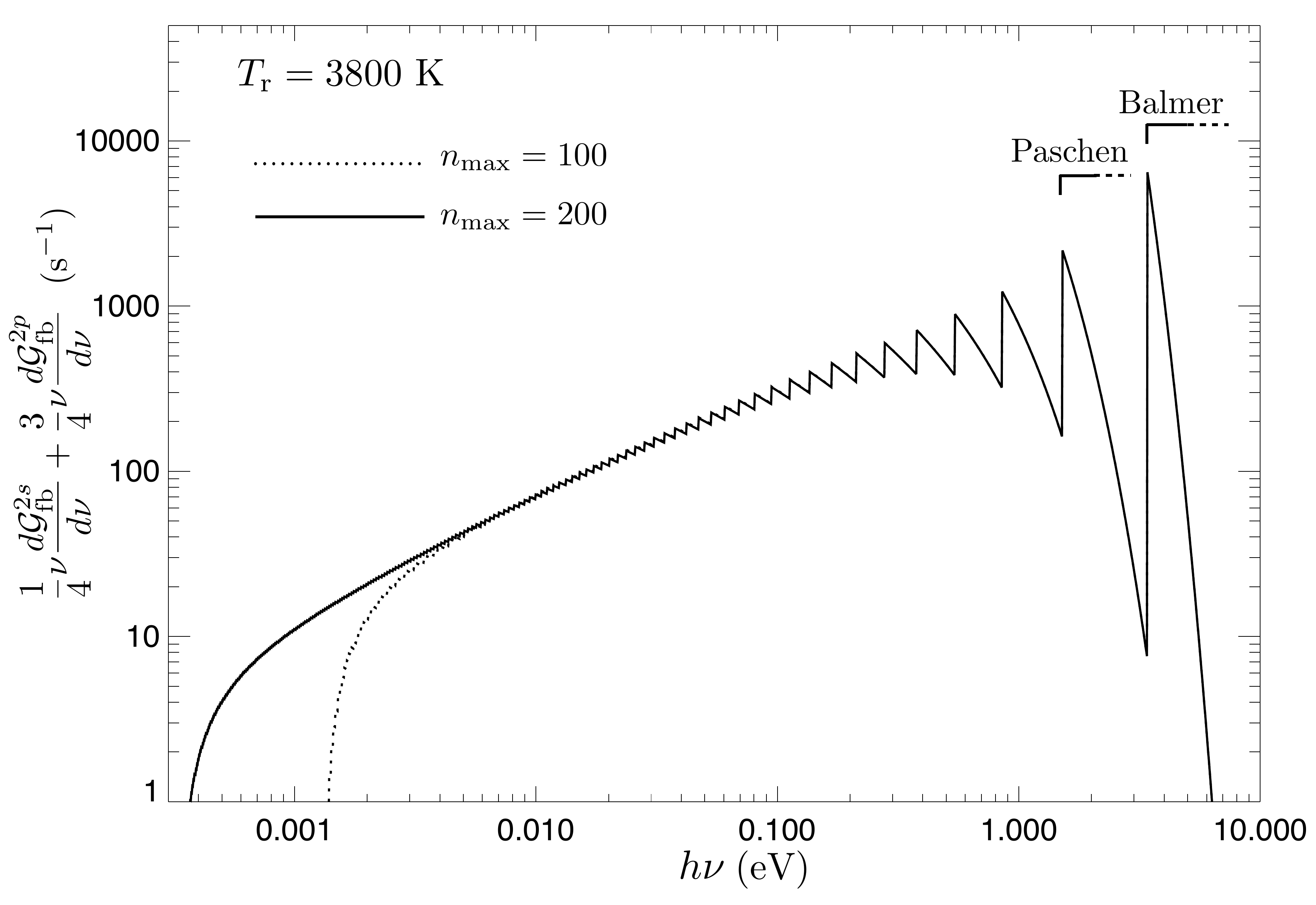}
\caption{\emph{Upper panel}: Weighted average of effective conductances $\frac14 \mathcal{G}_{n'n}^{2s} + \frac34
\mathcal{G}_{n'n}^{2p}$ at $T = 3800$ K, as a function of transition
energy $h \nu_{n'n}$,
for $n_{\max} = 100$ (open circles) and $n_{\max} = 200$ (dots). The nearly vertical families of
points correspond to a given series $n'\rightarrow n$ with fixed $n$,
such as the Balmer series $n'\rightarrow 2$, the Paschen series
$n'\rightarrow 3$, etc..., with $n$ increasing from right to left. The families of diagonal lines
correspond to a given order $(n + \Delta n) \rightarrow n$ with fixed
$\Delta n$, such as the $\alpha-$transitions $(n+1) \rightarrow n$, the
$\beta-$transitions $(n+2) \rightarrow n$ etc..., with lower
conductances for higher order transitions. \emph{Lower panel}: Weighted average of effective free-bound conductances per
  log-frequency interval $\frac14 \nu \frac{d\mathcal{G}_{\rm
      fb}^{2s}}{d \nu} + \frac34
\nu \frac{d\mathcal{G}_{\rm fb}^{2p}}{d \nu}$ at $T = 3800$ K, as a function of photon
energy $h \nu$, for $n_{\max} = 100$ (dotted line) and $n_{\max}
= 200$ (solid line). The edges correspond to photoionization
thresholds from various shells. We only show the effective
conductances for these two values of $n_{\max}$ in order not to
cluster the figure; proper convergence with $n_{\max}$ is discussed in
Section \ref{sec:results}.} \label{fig:conductances} 
\end{figure*}

We show some of the computed effective conductances in
Fig.~\ref{fig:conductances}, for a temperature $\Tr = 3800$ K
(corresponding roughly to the peak of the emission).

\subsection{Practical simplification}

Equation~(\ref{eq:master-eq}) is a rather remarkable result in its
raw form, but its implementation without further simplification could be somewhat
cumbersome: if considering excited states up to principal quantum
number $n_{\max}$, one would in principle need to tabulate of order
$n_{\max}^2/2$ functions $\mathcal{G}_{n'n}(\Tr)$ as a function of
temperature. For $n_{\max}$ of a few hundred, required for a fully
converged spectrum in the GHz region, one would need to store tens of
thousands of coefficients on a fine grid of temperature values,
interpolate them at each redshift, and compute the recombination
spectrum on a grid fine enough to resolve all the resonances. In
addition, whereas hydrogen and singly ionized helium benefit from the accidental
energy degeneracy between angular momentum substates, this is not the
case for neutral helium, which as a consequence has a much larger set of lines.

In order to save a significant amount of memory with negligible cost
in accuracy (as we shall demonstrate in the next section), we group resonances into bins
$b$ of finite width $\Delta \nu_b$, so we use
\beq
\frac{d\mathcal{G}^{e,i}}{d \nu}\Big{|}_{\rm used} = \sum_b
\mathcal{G}^{e,i}_b \delta(\nu - \nu_b),
\eeq 
where $\nu_b$ are the bin centers, and 
\beq
\mathcal{G}^{e,i}_b \equiv \sum_{n<n'} \mathcal{G}_{n'n}^{e,i}~
\mathbf{1}_b(\nu_{n'n}) + \Delta \nu_b \frac{d\mathcal{G}_{\rm fb}^{e,i}}{d \nu}(\nu_b),
\eeq
where $\mathbf{1}_b(\nu)$ is unity
if $\nu$ falls inside the bin $b$ and zero elsewhere. The
characteristic error resulting from this simplification should be of
the order of a few times $\Delta \log\nu_b$, the log-spacing between bins, since
the recombination timescale is a few times shorter than the Hubble
time\footnote{This can be understood from Eq.~(\ref{eq:redshift}): a fractional error
$\Delta \log \nu$ in the rest-frame frequency translates into the same fractional
error in the redshift of emission, hence a fractional error
$\Delta \log \nu/(H \tau)$ on the emissivity, where $\tau \sim H^{-1}/$few is the
characteristic time of evolution of the populations.}. In order to reduce the error induced by this simplification, we enforce that the
bin centers coincide with the lowest order transition they
may contain (and are logarithmically spaced otherwise). As can be seen
from Fig.~\ref{fig:conductances}, effective conductances indeed decrease with
increasing transition order (see also Table 1 of Ref.~\cite{Rubino_2006}).

With this discretization, the emissivity for bound-bound and
free-bound transitions between excited states becomes
\barr
&&\frac1{\nh}j_{\nu}\big{|}_{\rm used}^{\textrm{high-}n} = \frac{h \nu}{4 \pi}\times\sum_b \Big{\{}\mathcal{G}^e_b ~\nh x_e x_p \frac{\Delta
    T}{T} \nonumber\\
&&~~~~~~~~~~~~~~~~~~~~~~~~~~~ - \sum_{i = 2s, 2p}\mathcal{G}^{i}_b ~\Delta x_{i} \Big{\}} \delta(\nu - \nu_b).
 \earr

\subsection{Lyman-$\alpha$ and $2s-1s$ emission}

The net emissivity in the Lyman-$\alpha$ line is given by
\barr
\frac{j_{\nu}\big{|}_{\rm Ly \alpha}}{\nh} &=&  \frac{h \nu}{4 \pi} \left(x_{2p}
  - 3 x_{1s} \rme^{- h \nu_{\rm Ly \alpha}/k \Tr} \right) \nonumber\\
&\times& P_{\rm esc}
A_{2p,1s} \delta(\nu - \nu_{\rm Ly \alpha}), \label{eq:Lya}
\earr
where $P_{\rm esc}$ is the escape probability for the optically thick
Lyman-$\alpha$ line. In the Sobolev
approximation (see for example Ref.~\cite{AGH_2010} for a detailed derivation), it is given by
\beq
P_{\rm esc} = \frac{8 \pi H \nu^3}{3 c^3 \nh x_{1s} A_{2p,1s}}.\label{eq:Pesc}
\eeq

The $2s-1s$ two-photon emissivity is given by
\beq
\frac{j_{\nu}\big{|}_{2\gamma}}{\nh} = \frac{h \nu}{4 \pi} \frac{d
  \Lambda_{2s,1s}}{d \nu} \left[x_{2s}(1 + f_{\nu'})(1 + f_{\nu}) - x_{1s}f_{\nu'}
  f_{\nu} \right],
\eeq
where $\nu' = \nu_{\rm Ly \alpha} - \nu$ and $f_{\nu}, f_{\nu'}$ are
the values of the photon
occupation number at $\nu, \nu'$. This expression properly accounts for stimulated
decays and absorption of non-thermal photons (emitted in the
Ly$\alpha$ line for example). In order to be consistent with our
simple ``standard'' treatment of two-photon decays (see next section), we shall
neglect these two effects here and use the approximate expression
\beq
\frac{j_{\nu}\big{|}_{2\gamma}}{\nh} \approx \frac{h \nu}{4 \pi} \frac{d
  \Lambda_{2s,1s}}{d \nu} \left[x_{2s}- x_{1s}\rme^{- h \nu_{\rm Ly
      \alpha}/ k \Tr} \right]. \label{eq:2g}
\eeq
We approximated the differential two-photon decay rate $d \Lambda_{2s,1s}/d \nu$ with the fitting
formula of Ref.~\cite{Nussbaumer_1984}.

Note that neither Eq.~(\ref{eq:Lya}) nor Eq.~(\ref{eq:2g}) are
accurate at the percent level. If needed, it would be relatively
straightforward to include the appropriate corrections.

\subsection{Fast part of the computation} \label{sec:fast-part}

In this section we briefly recall how the recombination history can be
very efficiently computed with the EMLA method \cite{EMLA} and give explicit equations
for the ``voltages'' that source the emissivities.

The EMLA formulation of the problem collapses all the fast
and thermally-mediated interior transitions into the effective
recombination coefficients $\mathcal{A}_{i}(\Tm, \Tr)$,
photoionization rates $\mathcal{B}_i(\Tr)$ and transition
rates $\mathcal{R}_{i\rightarrow j}(\Tr)$ for $i,j = 2s, 2p$. All
the complication of the recombination computation then resides in the
slow transitions to the ground state, where a proper time-dependent
radiative transfer calculation is required for percent-level precision. Here we are not worried about such subtleties, and use the
following simple prescriptions for decay rates to the ground
state. For the two-photon transitions from $2s$, we neglect
stimulated decays and absorptions of non-thermal photons, so the net
decay rate is
\beq
\dot{x}_{1s}\big{|}_{2\gamma} = - \dot{x}_{2s}\big{|}_{2\gamma} =
\Lambda_{2s,1s}\left[x_{2s} - x_{1s}\rme^{- E_{21}/k \Tr}\right],
\eeq
where $\Lambda_{2s,1s} \approx 8.22$ s$^{-1}$ is the spontaneous
two-photon decay rate. For the net decay rate in the Lyman-$\alpha$
line, we use the Sobolev approximation,
\beq
\dot{x}_{1s}\big{|}_{\rm Ly \alpha} = - \dot{x}_{2p}\big{|}_{\rm Ly
  \alpha} =  R_{\rm Ly \alpha}\left[x_{2p} - 3 x_{1s}\rme^{- E_{21}/k \Tr}\right],
\eeq
where $R_{\rm Ly \alpha} \equiv A_{2p,1s} P_{\rm
  esc}$ is the net decay rate in the Ly-$\alpha$ line, accounting for the small escape probability
of resonant photon, given in Eq.~(\ref{eq:Pesc}).

The steady-state rate equations for $2s$ and $2p$ then read
\barr
0 \approx \dot{x}_{2s} &=& \nh x_e x_p \mathcal{A}_{2s} + x_{1s}
\rme^{- E_{21}/k \Tr}\Lambda_{2s,1s}\nonumber\\
 &+&x_{2p} \mathcal{R}_{2p
  \rightarrow 2s} - \Gamma_{2s} x_{2s},\\
0 \approx \dot{x}_{2s} &=& \nh x_e x_p \mathcal{A}_{2p} + 3 x_{1s}
\rme^{- E_{21}/k \Tr}R_{\rm Ly \alpha} \nonumber\\
&+&x_{2s} \mathcal{R}_{2s
  \rightarrow 2p} - \Gamma_{2p} x_{2p},
\earr
where the effective inverse lifetimes of the $2s$ and $2p$ states are
given by
\barr
\Gamma_{2s} &\equiv& \mathcal{B}_{2s} + \mathcal{R}_{2s \rightarrow
  2p} + \Lambda_{2s,1s}, \\
\Gamma_{2p} &\equiv& \mathcal{B}_{2p} + \mathcal{R}_{2p \rightarrow
  2s} + R_{\rm Ly \alpha}.
\earr
This simple 2 by 2 system (which is \emph{exact} in the limit
that $2s$ and $2p$ are the only interface states) can be solved
analytically. Using detailed balance relations satisfied by the
effective rates, we find that the departures from Saha equilibrium
are given by
\barr
\Delta x_{2s} &=& \frac{s_{2s}+ s_{2p} \frac{\mathcal{R}_{2p\rightarrow
    2s}}{\Gamma_{2p}}}{\Gamma_{2s} - \mathcal{R}_{2s \rightarrow 2p}\frac{\mathcal{R}_{2p\rightarrow
    2s}}{\Gamma_{2p}}}\\
\Delta x_{2p} &=&\frac{s_{2p}+ s_{2s} \frac{\mathcal{R}_{2s\rightarrow
    2p}}{\Gamma_{2s}}}{\Gamma_{2p} - \mathcal{R}_{2p \rightarrow 2s}\frac{\mathcal{R}_{2s\rightarrow
    2p}}{\Gamma_{2s}}}, 
\earr
where we have defined 
\barr
s_{2s} &\equiv& \nh x_e x_p ~\Delta \mathcal{A}_{2s} + \Delta
  x_{1s} ~ \rme^{- E_{21}/k \Tr}\Lambda_{2s,1s} ,\\
s_{2p} &\equiv& \nh x_e x_p ~\Delta \mathcal{A}_{2p} + 3~ \Delta
  x_{1s}~ \rme^{- E_{21}/k \Tr} R_{\rm Ly \alpha},
\earr
where $\Delta \mathcal{A}_i \equiv \mathcal{A}_i(\Tm, \Tr) -
\mathcal{A}_i(\Tr, \Tr)$ and $\Delta x_{1s}$ is the departure of the
ground state population from Saha equilibrium with the plasma, defined
as in Eq.~(\ref{eq:Dxnl-def}). Similar expressions can easily be obtained for
the departures from Boltzmann equilibrium with the ground state, needed for the
Lyman-$\alpha$ and $2s-1s$ emissivities.

The rate of change of the free-electron fraction
can be obtained, for example, from the departures from Saha
equilibrium computed above:
\beq
\dot{x}_e = -\sum_{i = 2s, 2p}\left[\nh x_e x_p \Delta \mathcal{A}_i -
  \Delta x_i \mathcal{B}_i \right].
\eeq

Finally, the matter temperature is obtained from
the Compton-heating equation, 
\beq
\dot{T}_{\rm m} = - 2 H \Tm + \frac{8 x_e
  \sigma_{\rm T} a_{\rm r} \Tr^4}{3(1 + x_e + f_{\rm He}) m_e c}(\Tr
- \Tm),
\eeq
where $\sigma_{\rm T}$ is the Thomson cross-section, $a_{\rm r}$ is the radiation
constant, $m_e$ is the electron mass and $f_{\rm He}$ is the He:H abundance
ratio.
At high redshifts where $\Tm$ is locked to $\Tr$, one can use the quasi-steady-state solution \cite{Hirata_2008}
\barr
\frac{\Delta T}{T} \approx  \frac{3 H (1 + x_e + f_{\rm He}) m_e c}{8 x_e
  \sigma_{\rm T} a_{\rm r} \Tr^4}.
\earr

As an aside, we point out that it is straightforward to include the effect of
dark matter annihilations in this system of equations (see for example Refs.~\cite{Chluba_2010, Giesen_2012}). One simply has to add an additional
photoionization term to the free-electron fraction evolution equation, a
heating term to the matter-temperature evolution equation, and properly account
for the additional excitations when solving for the populations of the
excited states.

\subsection{Numerical solution of the radiative transfer equation}

We first solve for the ionization history and matter temperature as described in Section \ref{sec:fast-part}, using
the code \textsc{HyRec} in \texttt{EMLA} mode, which we have adapted to
also extract the populations of the
excited states (more precisely, their departures
form equilibrium). We store these quantities on a fine redshift grid
ranging from $z = 2500$ to $z = 400$ for future interpolation.

We follow the spectral distortion from $z = 2500$ to $z = 400$ on a constant energy range
$10^{-5} \textrm{eV} < E < 10.2$ eV, where the upper limit corresponds
to the Lyman-$\alpha$ frequency. We assume a purely thermal spectrum
blueward of Ly-$\alpha$. Below $z = 400$, we set the emissivity to
zero and freely redshift the distortion down to $z = 0$.

We discretize the $2s-1s$ emissivity on the same grid as the high-$n$ transitions
emissivity. Our total discretized emissivity therefore has the form
\beq
j_{\nu}\big{|}_{\rm used} = \sum_b J_b \delta(\nu - \nu_b).
\eeq

At each time step, we first redshift the pre-existing spectral distortion from one bin to
the next lower one, using Eq.~(\ref{eq:free-redshifting}). We then
update it by adding the emission from the ``lines'' at frequencies
$\nu_b$, as in Eq.~(\ref{eq:jump}). This procedure was used (with
additional complications due to large optical depths and frequency diffusion) in
Refs.~\cite{Hirata_2008, Hyrec}. This method forces the timestep
$\Delta \log a$ to be no greater than the smallest bin separation
$\Delta \log \nu$.  

In our fiducial computation, we have grouped effective conductances in
bins of width $\Delta \log \nu = 10^{-2}$ (and we recall that the central
frequency assigned to each bin is chosen to coincide with the frequency of the lowest-order
line it contains), and used a time-step $\Delta \log a = 5 \times 10^{-3}$. We have checked that reducing the
bin width and timestep by a factor of 10 leads to maximum changes of at most a
percent over the whole range of frequencies considered, with a
root-mean-square difference of the order of 0.15\%. We also checked
that the spectrum is converged with respect to the redshift range over which it is computed -- this stems from the fact that the emissivities
are relatively sharply peaked around $z \sim 1300$, as we show in the
next section.

\subsection{Results}

All quantities shown in this section are evaluated for a flat universe
with fiducial cosmological parameters consistent with the latest WMAP
results \cite{WMAP}: $T_{\rm cmb} = 2.728$ K, $\Omega_b h^2 =
0.022$, $\Omega_m h^2 = 0.13$, $\Omega_{\Lambda} h^2 = 0.34$, $Y_{\rm
  He} = 0.24$, $N_{\nu, \rm eff} = 3.04$.

\begin{figure*}
\includegraphics[width = 89 mm]{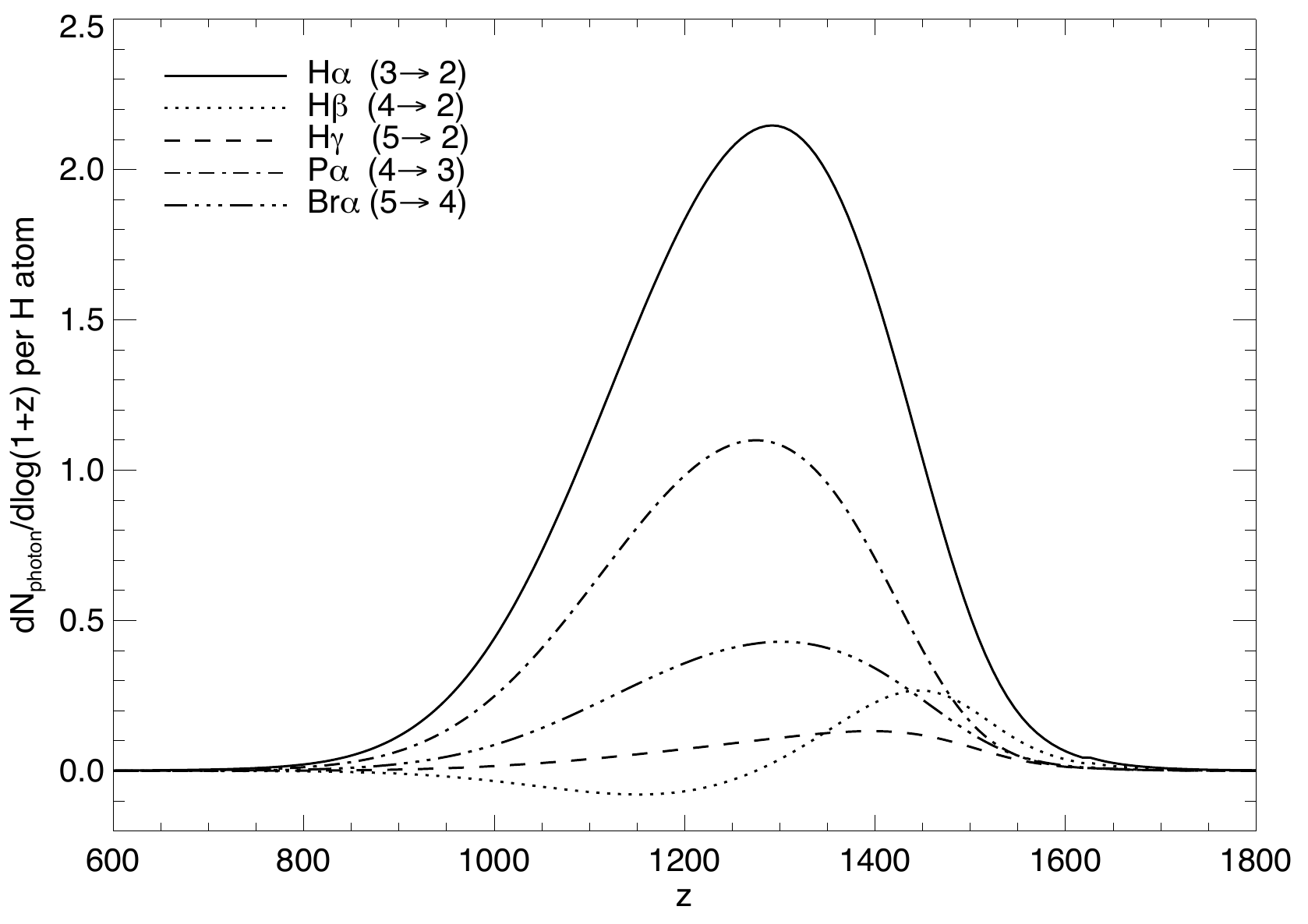}
\includegraphics[width = 89 mm]{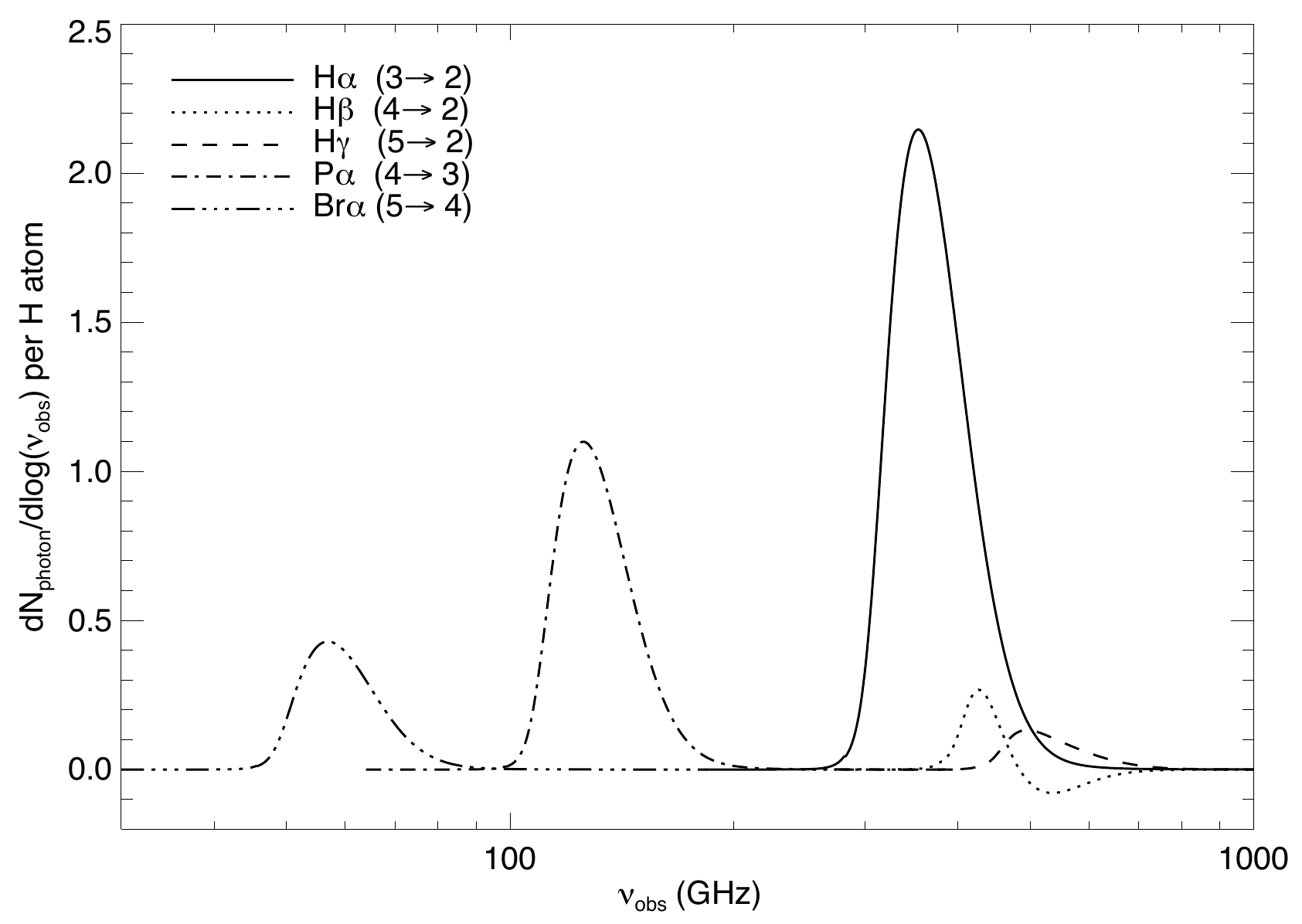}
\caption{Number of photons per hydrogen atom emitted per logarithmic
  redshift interval (or equivalently per logarithmic interval of
  today's observation frequency), for several bound-bound
  transitions. In the left panel, this quantity is plotted as a
  function of emission redshift. In the right panel, it is plotted as
  a function of observed frequency today. The two are related by
  $\nu_{\rm obs} = \nu_{\rm em}/(1 + z_{\rm em})$.} \label{fig:dNgamma} 
\end{figure*}

In Fig.~\ref{fig:dNgamma}, we show the number of photons emitted per
hydrogen atom per logarithmic redshift interval or per logarithmic
interval of observed frequency, for the
first few transitions of the Balmer series, and for the $\alpha$-transitions
of the first few series. We see that in general the number of emitted photons decreases
rapidly with the order of the transition within a series, and
decreases as well (but less rapidly) for higher series. Note that some transitions may show
absorption, as the H$\beta$ transition \cite{Rubino_2006}. We find
that a total of 0.63, 0.019, 0.036, 0.32 and 0.13 photons are emitted per hydrogen
atom in the H$\alpha$, H$\beta$, H$\gamma$, P$\alpha$ and
Br$\alpha$ transitions, respectively. 

\begin{figure*}
\includegraphics[width = 130 mm]{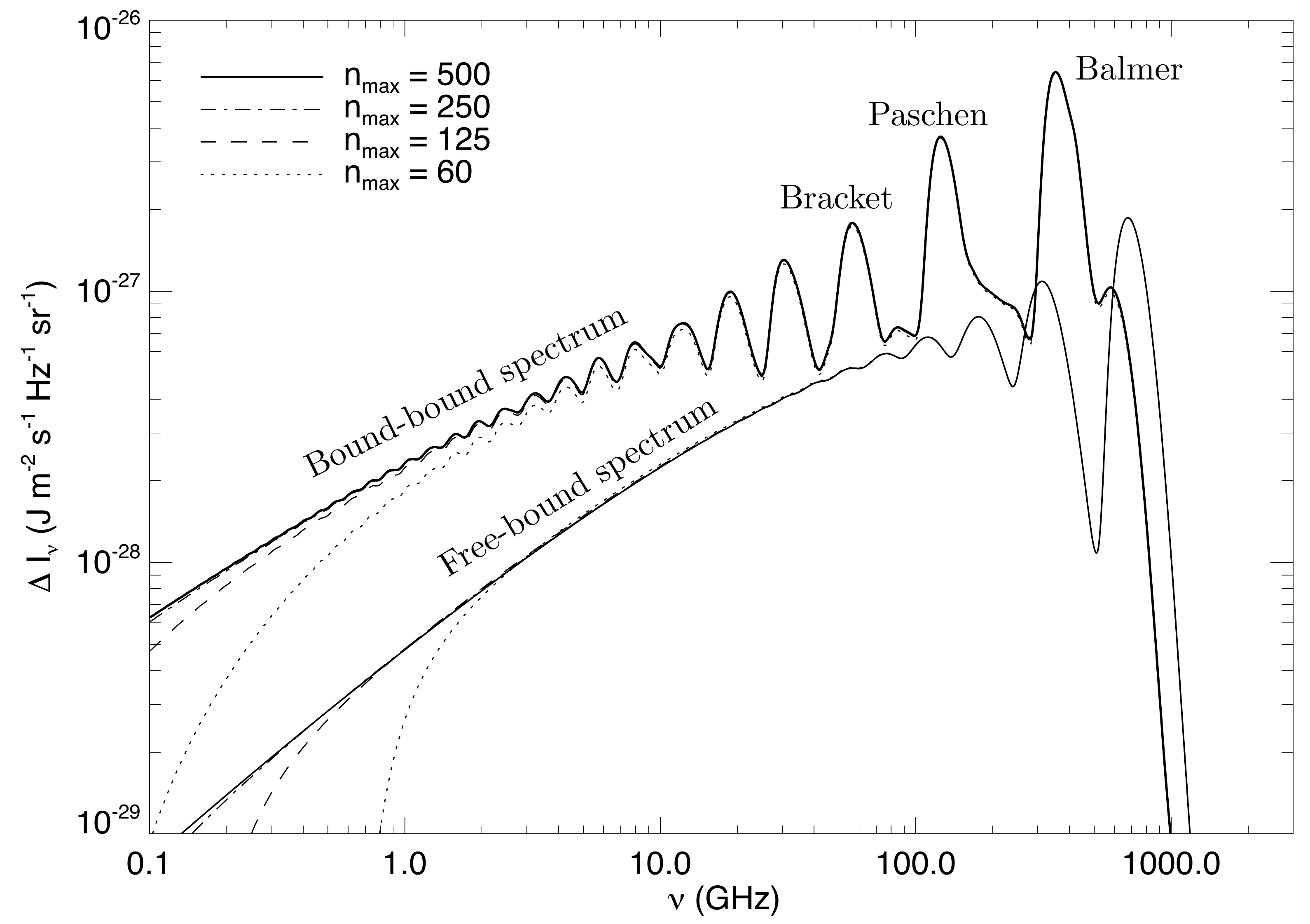}
\caption{Bound-bound (\emph{not} including $2s\rightarrow 1s$ and
  Ly-$\alpha$ photons) and
  free-bound spectra for various values of the cutoff principal
  quantum number $n_{\max}$.}\label{fig:nmax} 
\end{figure*}

Figure \ref{fig:nmax} shows the convergence of the
high-$n$ bound-bound and free-bound spectra with $n_{\max}$. We find that the fractional difference in the total spectrum between
  $n_{\max} = 250$ and $n_{\max} = 500$ is less than a percent for
  $\nu \gtrsim 0.5$ GHz. Nothing formally limits us from going beyond
  $n_{\max} = 500$ with our method, since the tabulation of effective
  conductances needs to be done only once. However, consistently
  computing the spectrum at low frequencies would also require
  accounting for free-free absorption and collisional transitions
  \cite{CRS_2007, Chluba_Vasil_2010}, which we do not include here. We
  therefore limit ourselves to $n_{\max} = 500$ for now, keeping in mind
  that the spectrum obtained is not accurate below a few tenths of GHz.

\begin{figure*}
\includegraphics[width = 130 mm]{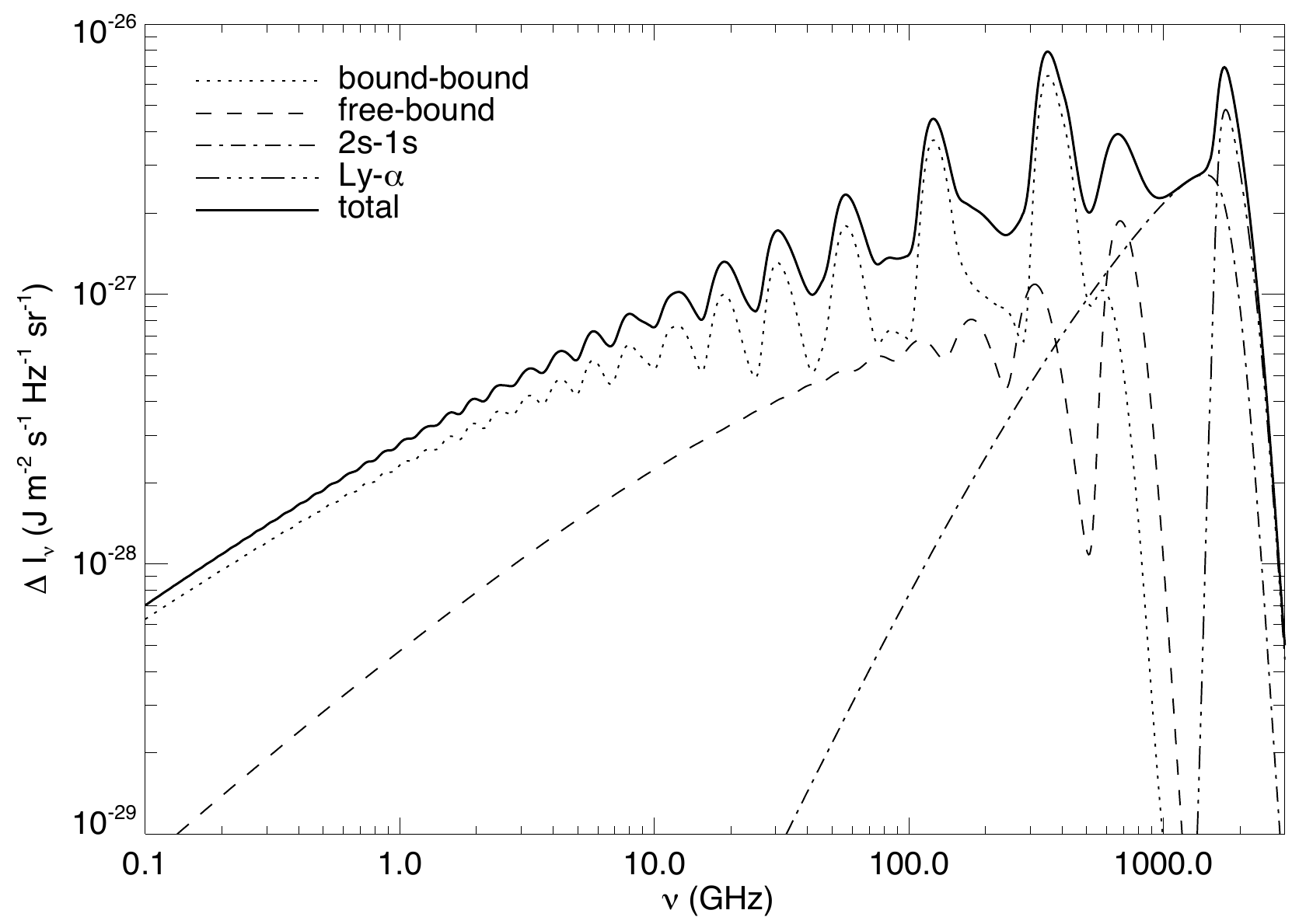}
\caption{Total spectral distortion
  created by hydrogen emission (neglecting the influence of helium),
  as well as individual contributions from various processes, using
  $n_{\max} = 500$. The ``bound-bound'' curve only accounts for $n' \rightarrow n\geq 2$
  transitions. We display $\Delta I_{\nu}$ in
the same units as Refs.~\cite{Rubino_2006, CS_2006} for an easier
comparison by eye. } \label{fig:spectrum} 
\end{figure*}

Figure \ref{fig:spectrum} shows the total recombination
spectrum today, as well as its subcomponents: bound-bound, free-bound,
two-photon emission from $2s-1s$ decays, and Lyman-$\alpha$
emission. Note that in the present work
we have not accounted for any of the radiative transfer effects
recently investigated with the purpose of obtaining high-accuracy
recombination histories (see for example Refs.~\cite{CS_2006_2s,
  Kholupenko_2006, Grachev_Dubrovich_2008,
  CS_2008_2g, Hirata_2008, CS_2009, Hirata_Forbes_2009, AGH_2010} and
many more references therein). It is to be expected that these
corrections will lead to a few percent corrections to the
recombination spectrum; nevertheless, we are far from even detecting
the recombination spectrum, and such refinements are not yet needed
for this purpose. The effective conductance method is, moreover,
oblivious to all the complications that may occur between interface states
and the ground state, and could be very easily adapted to include
these effects\footnote{A subtlety might arise when dealing with
  two-photon decays from higher levels, as one must avoid
  double-counting of the low-frequency photons.}. If needed for some
reason, a high-accuracy Lyman-lines and $2s-1s$ spectrum can be extracted from the
modern recombination codes \textsc{HyRec} \cite{Hyrec} and
\textsc{CosmoRec}\footnote{http://www.cita.utoronto.ca/$\sim$jchluba} \cite{Cosmorec}, that do account for these radiative
transfer effects.

Let us point out that even though we have performed all computations
with the correct matter temperature, we found that setting
$\Tm = \Tr$ (and doing so consistently, including when computing the
ionization history and departures from Saha equilibrium), leads to an
error of at most $0.4\%$ for $\nu \geq 0.1$ GHz. It would therefore be
sufficient, at the percent level of accuracy, to assume $\Tm = \Tr$
for the spectrum computation (this assumption is not valid if one
wishes to compute the low-redshift tail of the recombination history,
when the two temperatures may differ significantly). 

Finally, we have compared our results with those of
Ref.~\cite{CS_2006} and found a very good agreement. More thorough
comparisons will be made once we implement emission from helium as well.

The recurring time for computing a full recombination spectrum,
independent of $n_{\max}$ used for the effective conductances, is
approximately 0.1 second on a standard workstation. This is
four to six orders of magnitude faster than the standard multilevel
atom method\footnote{One can speed up the computation of the spectrum
  by a factor of $\sim 10$ if precomputing the recombination history
  with the EMLA method \cite{EMLA} and then computing the spectrum with the standard
  MLA method but with a larger timestep (J.~Chluba, private
  communication). This would still be a few orders of magnitude slower
than the method we present here.}, depending on the value used for $n_{\max}$ \cite{Chluba_Vasil_2010}.

\section{Conclusions and future directions} \label{sec:conclusion}

In this paper we have introduced a new and highly efficient method to compute the
primordial recombination spectrum. Our method relies on the
factorization of the problem in a computationally expensive but
cosmology-independent part and a very fast cosmology-dependent part --
the computational efficiency of the latter part is itself due to a similar method of factorization,
the effective multilevel atom method, introduced in an earlier work
\cite{EMLA}. The recurring cost of a spectrum computation with this method is a
fraction of a second, at least four orders of magnitude faster than
computations carried out with the standard multilevel atom method. 

The computations carried out here relied on several simplifying
assumptions. First, we have restricted ourselves to the ``standard'' (i.e. simplified) transitions
from $2s$ and $2p$ to the ground state, and have neglected a whole suite of radiative
transfer effects that were shown to be important for a highly accurate
recombination history and are efficiently computed by modern
recombination codes \cite{Hyrec, Cosmorec}. To our knowledge, no
computation of the primordial recombination spectrum thus far has also
included all these effects consistently. Including them with our formalism should not
raise any major issues, but they are expected to induce corrections to the spectrum at the level of a few percent at
most (though in principle one should check this statement precisely). The
second aspect we neglected is collisional transitions. They were shown
to modify the spectrum mostly in the low frequency regime $\nu \lesssim 0.1$
GHz, where the recombination distortions are very smooth and therefore
probably difficult to observe \cite{Chluba_Vasil_2010}. Finally, we
neglected broadening of the lines by electron scattering, as well as free-free
absorption. The former effect should increase the width of the broad
spectral features by less than a percent of their central frequency, and the latter
strongly affects the spectrum at frequencies below $\sim 0.1$ GHz
\cite{CRS_2007}. Since all these effects are small in the frequency
regime $\nu \gtrsim 0.1$ GHz, we do not consider them a priority before the observability
of the recombination spectrum is solidly established.

In this work we have only dealt with the line and continuum emission
from recombining hydrogen atoms, and have neglected the effect of
helium on the recombination spectrum (but we did account for helium
when computing the ionization history of the plasma). The presence of about 0.08 helium nuclei
per hydrogen nucleus will modify our results in the two following ways. First, photons emitted in the HeI
$2^1$P$-1^1$S line (the equivalent of the hydrogen Ly-$\alpha$ line)
with an energy of 21.2 eV can photoionize some of the few neutral
hydrogen atoms already present at the epoch of HeII$\rightarrow$ HeI
recombination \cite{Hu_Scott_1995, Kholupenko_2007, Switzer_2008_1, CS_2010_helium_feedback}. This effect is
already included in our code when computing the ionization fraction of
helium \cite{Hyrec}, but we have not accounted here for the small
departure from Saha equilibrium that it induces for hydrogen, and the
resulting line emission. This process was studied in detail in
Ref.~\cite{CS_2010_helium_feedback}, who showed that it leads to
pre-recombination emission features in the hydrogen spectrum. Inclusion of this effect with our method presents no additional
complication: one only needs to properly compute the small departures
from equilibrium in hydrogen, given the helium ionizing photons. The
second and more direct consequence of having helium nuclei is that
they too recombined, and in the process, emitted a few nonthermal
photons per helium nucleus. Because helium recombined in two stages (at redshifts $z \sim 6000$ for HeIII$\rightarrow$
HeII recombination and $z \sim 2000$ for HeII$\rightarrow$ HeI
recombination), this results in $\sim 0.16$ helium recombination
photons per hydrogen recombination photon, which is a non-negligible
contribution to the spectrum, and can be used to probe the primordial
helium abundance before the first stars formed \cite{Rubino_2008}. Computing the emission from helium
recombination can be achieved with the exact same method, the only
additional work would be to compute effective conductances for HeI
(those for HeII can be easily obtained from the ones computed for
hydrogen by simple rescalings). Such additions are essential and we
shall include them in the near future.

An additional aspect that we have not treated here is the processing
of pre-existing spectral distortions by the recombining atoms. These
can lead to large enhancements of the line emission, even if the
initial smooth spectral distortion is tiny \cite{CS_2009_features}. Formally, one can generalize our method to
account for non-thermally mediated transitions, as long as the
underlying spectrum can be described by a small set of parameters. We
defer such a study to future work. Once the latter two aspects are implemented, we shall release a public
code to efficiently compute the recombination spectrum.

Besides having a certain aesthetic appeal, we believe that the factorization method presented
here will prove very useful in order to quantitatively study the
information content of the recombination spectrum. Most of our
knowledge about the early universe currently comes from the study of
\emph{spatial} variations of the thermal photon background, through
CMB anisotropy observations. An exciting
avenue is to probe the \emph{spectral} variations of the
CMB, which might inform us about the universe before the last-scattering redshift. A fast and accurate code to compute the primordial spectrum
will be of great help to study precisely if, what and how new information can be gained
from the primordial spectrum.

\section*{Acknowledgments}

I thank Chris Hirata for pointing out the analogy of the system of fast high-$n$
transitions with a circuit. I am grateful to Jens Chluba for providing recombination spectra for comparison
with the results presented here and to Matias Zaldarriaga for discussions
on this work. I am indebted to Simeon Bird, Jens
Chluba and Daniel Grin for reading the draft of this manuscript and
providing many useful comments. This work was completed at the Institute for Advanced Study with the support of the National Science Foundation grant
number AST-0807444 and the Frank and Peggy Taplin Membership. I also acknowledge support from Rashid Sunyaev at the Max Planck
Institute for Astrophysics during part of summer of 2012, when the first steps of this study were taken.






\bibliography{recomb_lines}

\end{document}